\documentclass[aps,preprint,amsmath,amssymb]{revtex4}
\usepackage{graphicx}
\usepackage{color}
\def\be{\begin{eqnarray}}
\def\en{\end{eqnarray}}

\def\g5{\gamma_{5}}
\def\ga{\gamma}

\def\e{\epsilon}

\def\be{\begin{eqnarray}}
\def\ed{\end{eqnarray}}

\def\non{\nonumber}
\def\lam{\lambda}
\def\UP{\cal U}
\def\la{\langle}
\def\ra{\rangle}

\begin{document}
\title{ \Large \bf
Unparticle phase effects}
\date{\today}

\author{ \bf  Chuan-Hung Chen$^{1,2}$\footnote{Email:
physchen@mail.ncku.edu.tw} and Chao-Qiang
Geng$^{3,4}$\footnote{Email: geng@phys.nthu.edu.tw}
 }

\affiliation{ $^{1}$Department of Physics, National Cheng-Kung
University, Tainan 701, Taiwan \\
$^{2}$National Center for Theoretical Sciences, Taiwan\\
$^{3}$Department of Physics, National Tsing-Hua University, Hsinchu
300, Taiwan  \\
$^{4}$Theory Group, TRIUMF, 4004 Wesbrook Mall, Vancouver, B.C V6T
2A3, Canada
 }

\begin{abstract}
Unparticles proposed by Georgi carry CP conserving phases in their
propagators. We demonstrate that these peculiar phases have an
important impact on CP violation. Without including the strong QCD
phases, we study the unparticle phase effects on the direct CP
asymmetries in the exclusive decays of $\bar B_d\to \pi^{+} \pi^{-}$
and $B\to \pi K$, in which the flavor changing neutral currents are
forbidden at tree level but induced by one-loop diagrams.
Interesting and consistent results comparing to the data are
obtained. In addition, we find that unparticles will significantly
enhance the differential branching ratio of  $b\to s \ell^{+}
\ell^{-}$ at the small invariant mass of $\ell^{+} \ell^{-}$. The
forward-backward asymmetries for $b\to s \ell^{+} \ell^{-}$ due to
unparticles are also explored.

\end{abstract}
\maketitle

\section{Introduction}
In Refs. \cite{Georgi1,Georgi2} Georgi has suggested that a scale
invariant sector with a non-trivial IR fixed point decoupled at a
large scale is associated with unparticles, which could couple to
the standard model (SM) particles at the TeV scale. Consequently,
the unparticle physics phenomenology have been extensively explored
in Refs.
\cite{Georgi1,Georgi2,TC,Luo,CH,Yan,Liao,Aliev,Catterall,Li,Lu,Steph,Fox,Greiner,Davou,Choudhury,XG,Aliev2,MR,Zhou,DY}.
Moreover, Georgi in Ref. \cite{Georgi2} has pointed out that the
unparticle propagators in the time-like region are associated with
some peculiar CP conserving phases
 depending on the non-integral number of the scale dimension $d_{\UP}$.
He has shown that these phases can induce some
 unusual CP conserving interference effects
between the time-like unparticle exchange
amplitudes and the SM amplitudes
in $e^{+}e^{-}\to\mu^{+}\mu^{-}$.
 The effect of the virtual unparticle propagation  has also been noticed in Ref. \cite{TC}.

 Recently, in Ref. \cite{CH} we have demonstrated that
 the peculiar CP conserving
phases in the unparticle propagators
can also play very important roles
on CP violation. We have explicitly examined the phase effects on the
direct CP asymmetries (CPAs) in $B_{d}\to \pi^{-}\pi^{+}$ and $B_{d}\to
\ell^{-}\ell^{+}$ decays based on
operators with fermion flavor changing neutral currents (FCNCs) at tree level. We have found that the direct CPAs in both decays could be large.
In this paper, we will examine the unparticle phase effects
on CP violation with only flavor conserving operators at tree level.
Specifically, we only consider those effects in which FCNCs
are forbidden at tree level like the SM but they can be generated by
one-loop diagrams including the penguin unparticle ones.

It is well known that in a decay process the direct CPA (${\cal
A}_{CP}$) depends on two types of phases, called weak ($\delta$) and
strong ($\phi$) phases which are CP violating and conserving,
respectively. In particular, one has that \be {\cal A}_{CP}&\propto&
\sin\delta\sin\phi\,. \ed Clearly, to have a sizable value of ${\cal
A}_{CP}$, both phases have to be nonzero and large. In the SM, the
weak CP violating phase is the unique phase
 in the $3\times 3$ Cabibbo-Kobayashi-Maskawa (CKM) quark mixing matrix
\cite{CKM}, which has been fixed by experiments \cite{Wolfenstein,PDG06}.
The CP conserving strong phase is process dependent, which is
normally hard to be determined due to hadronic uncertainties.
Since the unparticle phases appear in the propagators and conserve CP,
it is interesting to speculate that these phases could act as the
strong phases in some physical processes \cite{CH}.
To explore this possibility, we will concentrate on
B decays as there are many experimental CP violating phenomena \cite{PDG06} from the current B factories as well as future super-B facilities.
In particular, we will investigate direct CPAs in the decays of
$B\to\pi\pi$ and $B\to\pi K$.
It is clear that our study can be extended to other processes
such as $K$ and $D$ decays.

The  paper is organized as followed. In Sec. II, we present FCNCs
which are induced by the unparticle penguin diagrams. We show the
unparticle effects on charmless nonleptonic and semileptonic B
decays in Sec. III. We give our numerical analysis in Sec. IV and
conclude
 our results in Sec. V.

\section{ Flavor changing neutral currents induced by unparticle penguin diagrams} \label{sec:loop}

To study the low energy effects of unparticle physics, for
simplicity, we assume that unparticles only couple to the flavor
conserving fermion currents,
described by \cite{Georgi1,Georgi2}
 \be
   \frac{1}{\Lambda^{d_{\UP}-1}_{\UP}}\bar f \ga_{\mu} \left( C^{\rm f}_{L} P_L
   +C^{\rm f}_{R} P_R\right) f O^{\mu}_{\UP} \label{eq:lang_UP}
 \ed
where $P_{L(R)}=(1\mp \ga_5)/2$ and $O^{\mu}_{\UP}$ is the vector
unparticle operator. Clearly, at tree level, the fermion flavor is
conserved. Similar to the SM, FCNCs such as $f\to f'\, \UP $ can be
induced by the charged weak currents at one-loop level. Due to the
CKM mixing matrix element $V_{tb}\approx 1$ and the heavy top-quark
enhancement, we will concentrate on $B$ decays to illustrate some
important physics phenomena involving unparticles. Our discussions
can be straightforwardly generalized to $K$ and $D$ decays.
%%%%%%%%%%%%%%%%%%%%%%%%%%%%%%%%%%%%%%%%%%%%%%%%%%%%%%%%%%%%%
\begin{figure}[hpbt]
\includegraphics*[width=1.8 in]{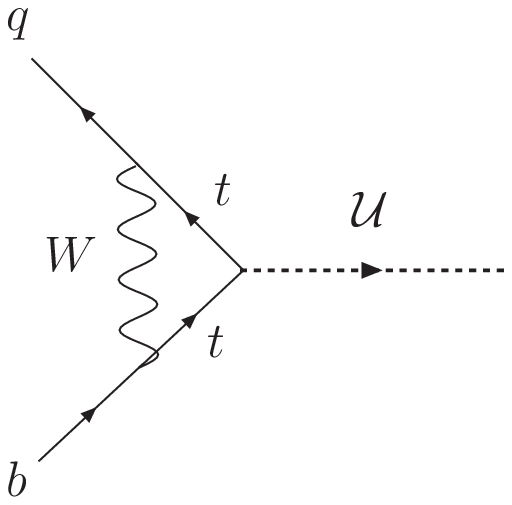}
\caption{Feynman diagram for $b\to q \UP$.}
 \label{fig:up}
\end{figure}
%%%%%%%%%%%%%%%%%%%%%%%%%%%%%%%%%%%%%%%%%%%%%%%%%%%%%%%%%%%%
%
The leading effective interaction for the  flavor changing transition of $b\to q$ is
induced by the unparticle penguin diagram shown in
Fig.~\ref{fig:up}, which leads to
 \be
{\cal L}_{\UP}= \frac{g^2}{\Lambda^{d_{\UP}-1}_{\UP}}V_{tb}
V^{*}_{tq}C^{qb} _{L} \bar{q} \ga_{\mu}  P_L b\, O^{\mu}_{\UP} \,,
 \label{eq:efflang}
 \ed
where
% \be
%  C^{qb}_{L}&=& \frac{C^{t} _{L}}{(4\pi)^2} I(x_t )  \,,\non\\
% I(x_t)&=&\frac{x_t(2+x_t)}{2(1-x_t)^2} \left( -1+x_t -\ln x_t \right)\,,
% \ed
  \be
  C^{qb}_{L}&=& \frac{1}{(4\pi)^2} I(x_t )  \,,\non\\
 I(x_t)&=&\frac{x_t(2 C^{t}_R +C^{t} _{L} x_t)}{2(1-x_t)^2} \left( -1+x_t -\ln x_t \right)\,,
 \ed
with $x_t=m^2_t/m^2_W$. Here, we have adopted the Feynman-'t Hooft
gauge and the contributions from the charged Goldstone boson have
been included. For simplicity, in the following analysis, we will
set $C^{t}_{R}\approx C^t_{L}$.
%Although the  couplings of unparticles to the t-quark
%contain
% the right-handed
%component, due to the W-boson interacting with the
%left-handed
%states, Eq.~(\ref{eq:efflang}) is only associated with $C^{t}_{L}$.

To obtain the unparticle-mediated effects, we need to know
the unparticle propagator, which
 is given by \cite{Georgi1,Georgi2}
  \be
  \int d^4x e^{i p\cdot x} \la 0| T\left( O^{\mu}_{\UP}(x)
O^{\nu}_{\UP}(0)\right) |0 \ra
 = i\Delta_{\UP}(p^2)\,e^{-i\phi_{\UP}}
%\left(-g^{\mu \nu} + \frac{p^{\mu} p^{\nu}}{p^2} \right)e^{-i
%\pi(d_{\UP}-2)}
\label{eq:uprop}
 \ed
where
 \be
    \Delta_{\UP}(p^2)&=& \frac{A_{d_{\UP}}}{2\sin(d_{\UP}
\pi)} \frac{ -g^{\mu \nu} + p^{\mu} p^{\nu}/p^2 }{\left(p^2
+i\e\right)^{2-d_{\UP}}}\,,
\non\\
\phi_{\UP}&=&(d_{\UP}-2)\pi \,,
 \ed
with
 \be%
A_{d_{\UP}}&=&  \frac{16 \pi^{5/2}}{(2\pi)^{2d_{\UP}}}
\frac{\Gamma(d_{\UP}+1/2)}{\Gamma(d_{\UP}-1) \Gamma(2 d_{\UP})}\,.
\label{eq:pro_up}
\ed%
Note that in Eq. (\ref{eq:uprop}) the phase factor arises from
$(-1)^{d_{\UP}-2}=e^{-i\pi(d_{\UP}-2)}$ and the vector operator is
assumed to satisfy the transverse condition $\partial_{\mu}
O^{\mu}_{\UP}=0$.  In terms of the effective interaction in
Eq.~(\ref{eq:efflang}) and the unparticle propagator in Eq.
(\ref{eq:uprop}), the effective Hamiltonian for $b\to q f\, \bar f$
can be written as
 \be
{\cal H}_{\UP}&=& - \frac{G_F}{\sqrt{2}}V_{tb} V^{*}_{tq}
\tilde{\Delta}_{\UP}(p^2) e^{-i\phi_{\UP}}\, \bar q \ga_{\mu} P_{L}
b\; \bar f\, \ga^{\mu}\left( C^{\rm f}_{L} P_L + C^{\rm f}_{R} P_R
\right)\, f\,, \label{eq:h_U}
 \ed
where $G_F=8\sqrt{2} g^2/m^2_W$ is the Fermi constant and
 \be
\tilde{\Delta}_{\UP}(p^2)=8 C^{qb}_{L} \frac{A_{d_{\UP}}}{2\sin
d_{\UP}\pi} \frac{ m^2_{W}}{p^2}
\left(\frac{p^2}{\Lambda^2_{\UP}}\right)^{d_{\UP}-1}\,.
 \ed
We note that $f$ can be neutrinos or charged leptons or quarks. We
remark that replacing $b$ by $s$ in
the effective Hamiltonian of Eq.~(\ref{eq:h_U}),
we may study
$s\to q f\bar f$ decays.

\section{ Charmless nonleptonic and semileptonic B
decays}\label{sec:decays}

For the nonleptonic decays of $b\to q q'' \bar{q''}$, we start with
the explicit expression of the effective Hamiltonian in the SM
\cite{BBL}
\begin{eqnarray}
 H_{{\rm eff}}&=&{\frac{G_{F}}{\sqrt{2}}}\sum_{q'=u,c}\xi^q_{q'}\left[
C_{1}(\mu) O_{1}^{(q)}(\mu )+C_{2}(\mu )O_{2}^{(q)}(\mu
)+\sum_{i=3}^{10}C_{i}(\mu) O_{i}(\mu )\right] \;,
\label{eq:hamiltonian}
\end{eqnarray}
where $\xi^q_{q'}=V_{q'b}V_{q'q}^{*}$ denotes the product of the CKM
matrix elements  and the operators $O_{1}$-$O_{10}$ are defined by
\begin{eqnarray}
&&O_{1}^{(q)}=(\bar{q}_{\alpha}q'_{\beta})_{V-A}(\bar{q'}_{\beta}b_{\alpha})_{V-A}\;,\;\;\;\;\;
\;\;\;O_{2}^{(q)}=(\bar{q}_{\alpha}q'_{\alpha})_{V-A}(\bar{q'}_{\beta}b_{\beta})_{V-A}\;,
\nonumber \\
&&O_{3}=(\bar{q}_{\alpha}b_{\alpha})_{V-A}\sum_{q''}(\bar{q''}_{\beta}q''_{\beta})_{V-A}\;,\;\;\;
\;O_{4}=(\bar{q}_{\alpha}b_{\beta})_{V-A}\sum_{q''}(\bar{q''}_{\beta}q''_{\alpha})_{V-A}\;,
\nonumber \\
&&O_{5}=(\bar{q}_{\alpha}b_{\alpha})_{V-A}\sum_{q''}(\bar{q''}_{\beta}q''_{\beta})_{V+A}\;,\;\;\;
\;O_{6}=(\bar{q}_{\alpha}b_{\beta})_{V-A}\sum_{q''}(\bar{q''}_{\beta}q''_{\alpha})_{V+A}\;,
\nonumber \\
&&O_{7}=\frac{3}{2}(\bar{q}_{\alpha}b_{\alpha})_{V-A}\sum_{q''}e_{q''} (\bar{q''}%
_{\beta}q''_{\beta})_{V+A}\;,\;\;O_{8}=\frac{3}{2}(\bar{q}_{\alpha}b_{\beta})_{V-A}
\sum_{q''}e_{q''}(\bar{q''}_{\beta}q''_{\alpha})_{V+A}\;,  \nonumber \\
&&O_{9}=\frac{3}{2}(\bar{q}_{\alpha}b_{\alpha})_{V-A}\sum_{q''}e_{q''} (\bar{q''}%
_{\beta}q''_{\beta})_{V-A}\;,\;\;O_{10}=\frac{3}{2}(\bar{q}_{\alpha}b_{\beta})_{V-A}
\sum_{q''}e_{q''}(\bar{q''}_{\beta}q''_{\alpha})_{V-A}\;,
\label{eq:operators}
\end{eqnarray}
with $\alpha$ and $\beta$ being the color indices, $C_{1}$-$C_{10}$
the
Wilson coefficients (WCs), $e_{q''}$ the
electric charge of $q''$ and $(\bar q'' q)_{V\pm A}=\bar
q''\ga^{\mu} (1\pm \ga_5) q''$. In Eq. (\ref{eq:hamiltonian}),
$O_{1}$-$O_{2}$ are from the tree level of weak interactions,
$O_{3}$-$O_{6}$ are the so-called gluon penguin operators and
$O_{7}$-$O_{10}$ are the electroweak penguin operators. Using the
unitarity condition, the CKM matrix elements for the penguin
operators $O_{3}$-$O_{10}$ can also be related by
\begin{eqnarray}
\xi^{q}_{u}+\xi^{q}_{c}=-\xi^{q}_{t}.
\end{eqnarray}

Comparing to Eq.~(\ref{eq:h_U}), we clearly see that the structures
of four-Fermi interactions with unparticle contributions are the same
as those of $O_3$ and $O_5$. Consequently, we can easily get
the  unparticle
contributions by replacing  $C_3$ and $C_5$ in the SM with
 \be
  C^{q\UP}_3(p^2) = C_3 + \frac{1}{4}\tilde{\Delta}_{\UP}(p^2) C^{ q}_L e^{-i\phi_{\UP}}  \,, \ \ \
  C^{q \UP}_5(p^2) = C_5 + \frac{1}{4}\tilde{\Delta}_{\UP}(p^2)C^{ q}_R
  e^{-i\phi_{\UP}}\,.
  \label{eq:c35}
   \ed
Accordingly,  the associated effective WCs could be
classified and re-expressed to be more useful forms by
\begin{eqnarray}
  a_{1} &=& C_{2}+\frac{C_1}{N_c}, \ \ \ a_{2}= C_1+
      \frac{C_2}{N_c}, \non \\
     a^{q\UP}_3&=&C^{q\UP}_{3}+\frac{C_{4}}{N_c}+ \frac{3}{2} e_{q} \left(C_9
     +\frac{C_{10}}{N_c} \right), \ \ \ a^{q\UP}_4 =C_{4}+\frac{C^{q\UP}_{3}}{N_c}+
     \frac{3}{2} e_{q} \left(C_{10} +\frac{C_{9}}{N_c} \right),\non \\
a^{q\UP}_5&=&C^{q\UP}_{5}+\frac{C_{6}}{N_c}+ \frac{3}{2} e_{q}
          \left(C_7+\frac{C_{8}}{N_c} \right),\ \ \
                  a^{q\UP}_6=C_{6}+\frac{C^{q\UP}_{5}}{N_c}+ \frac{3}{2} e_{q}
                  \left(C_8+\frac{C_{7}}{N_c} \right) \label{eq:effwcs}
\end{eqnarray}
where $N_c=3$ is the number of color.

\subsection{$B\to \pi \pi$  decays }

In this section, we are going to study the decays of
$B\to \pi \pi$ dictated by $b\to
d q\bar q$.
%,
Using the effective
operators displayed in Eqs.~(\ref{eq:operators}) and
(\ref{eq:effwcs}), it is easy to see that
the decays are tree dominated. However, it is clear that the
unparticle effects on $B^{-}\to\pi^{-} \pi^{0}$ could be neglected
if  $C^{d}_{L(R)} \approx C^{u}_{L(R)}$ as the effects are always
related to $-(a^{u\UP}_3 - a^{d\UP}_3) + (a^{u\UP}_5 - a^{d\UP}_5)$,
where the minus sign inside brackets is from the pion flavor wave
function, $|\pi^0 \ra = (\bar u u -\bar d d)/\sqrt{2}$,  and the
other one outside the brackets is due to the pseudoscalar decay
constant, defined by $\la P(p) | \bar f' \ga^{\mu} (1 \pm \ga_5) f
|0\ra = \pm i f_{P} p^{\mu}$. On the other hand, as no significant
CPA for $B^{-}\to \pi^{-} \pi^0$ is found based on the current
experimental world average, it should be a good scenario to take
$C^{d}_{L(R)} \approx C^{u}_{L(R)}$. Unfortunately, since the
unparticle contributions arise at one-loop level, we don't expect
that we can solve the problem of the large branching ratio (BR) on
$B_d \to \pi^0 \pi^0$, in which the tree contribution plays a
dominant role. Hence, we will concentrate on the CPA of $B_d \to
\pi^{+} \pi^{-}$.

It is known that the penguin effects on  $B_d\to \pi^{-} \pi^{+}$ are
significant even though  the decay is  tree dominated,
 In order to generate strong phases for the
CPA, in the SM the annihilation topology from $O_{6}\varpropto
(V-A)\otimes (V+A)$ plays a very important role. With the Fierz
transformation, since the corresponding QCD effects involve the
timelike form factor denoted by $\la \pi \pi| \bar q' (1+\ga_5)q
|0\ra$, the theoretical calculations are very uncertain. For
instance, with the QCD factorization \cite{QCDF}, to cure
divergences one needs to introduce free parameters to parametrize
the corresponding form factors. With the perturbative QCD approach
\cite{PQCD}, although singularities could be removed by transverse
degrees of freedom, the dominant dynamical scale is close to the
nonperturbative scale which is around $1$ GeV \cite{Chen}. Using the
soft-collinear effective theory, it is found that at the lowest
order in $\alpha_s$, the annihilation contributions are real
\cite{SCET}. In addition, these timelike form factors are all power
suppressed in $m_P/m_B$ with $m_P$ being the mass of the light
pseudoscalar \cite{CG}. Hence, it still needs to make lots of
efforts to fix the strong phases induced by the QCD effects.

\begin{figure}[hpbt]
\includegraphics*[width=4.5 in]{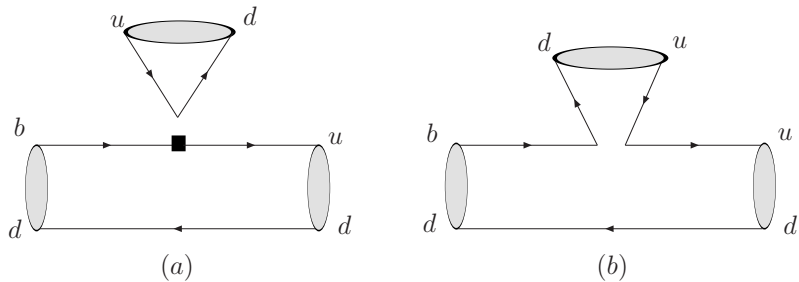}
\caption{Flavor diagrams from (a) tree and (b) penguin for $B_d\to
\pi^{-} \pi^{+}$ decay, where the square symbol denotes the weak vertex. }
 \label{fig:bpipi}
\end{figure}

As stated before, the CP-conserved phase in the unparticle
propagator
 could provide a
kind of strong phase needed for the CPA \cite{CH}. It has been
realized that the phase could contribute to the CPA of $B_d\to \pi^-
\pi^{+}$ with tree allowed FCNCs \cite{CH}. In this study, we take
the fermion flavor conservation at tree level like the SM. To
examine the influence of the
%peculiar
unparticle phase alone, we will neglect the uncertain strong phases induced
by QCD interactions.
By following the effective Hamiltonian in
Eq.~(\ref{eq:hamiltonian}), we present the decay topologies
in Fig.~\ref{fig:bpipi} where
%Fig.~\ref{fig:bpipi}
(a)[(b)] denotes the tree (loop) effects.
%%%%%%%%%%%%%%%%%%%%%%%%%%%%%%%%%%%%%%%%%%%%%%%%%%%%%%%%%%%%%
%%%%%%%%%%%%%%%%%%%%%%%%%%%%%%%%%%%%%%%%%%%%%%%%%%%%%%%%%%%%
Since $B_d\to \pi^- \pi^{+}$ is a color-allowed process, the
factorization
assumption is good enough to estimate the transition matrix
elements. Consequently, the decay amplitude for $\bar B_d\to \pi^{+}
\pi^{-}$ is
 given by
 \be
{\cal M}^{\UP}_{\pi^{+} \pi^{-}}&=& \frac{G_F}{\sqrt{2}} f_{\pi}
m^2_{B} F^{B\pi}_{0}(m^2_{\pi})\left[-V_{tb} V^{*}_{td} \left(
a^{u\UP}_4 + 2 r_{\pi} a^{u\UP}_{6}\right)+ V_{ub} V^{*}_{ud} a_1
\right]\,, \label{eq:amp_cpicpi}
%
%2{\cal M}^{\UP}_{\pi^{0} \pi^{0}}&=& \frac{G_F}{\sqrt{2}} f_{\pi}
%m^2_{B} F^{B\pi}_{0}(m^2_{\pi})\left[-V_{tb} V^{*}_{td} \left(
%a^{\UP}_4 + 2 r_{\pi} a^{\UP}_{6}\right)- V_{ub} V^{*}_{ud} a_2
%\right]\,.
 \ed
where the form factor $F^{B\pi}_0$ is defined by
 \be
\la \pi(p) | \bar u \ga_{\mu} b|\bar B(p_B)\ra&=&
\left[(p_B+p)_{\mu}-{m^2_B\over q^2}\,
q_{\mu}\right]F^{B\pi}_{1}(q^2)+ {m^2_B\over q^2}\, q_{\mu}
F^{B\pi}_0(q^2)
 \ed
 with $q=p_B-p$ and
$r_{\pi}=m^{0}_{\pi}/m_{B}$ associated with $\langle \pi| \bar{d}
\gamma_{5} u | 0\rangle = i f_{\pi} m^{0}_{\pi}$. For the light
meson production in $B$ decays, we take $F^{B\pi}(m^2_{\pi})\approx
F^{B\pi}_{0}(0)$. Hereafter, we will use $F^{B\pi}_0$ instead of
$F^{B\pi}_{0}(0)$. As a result, the BR and CPA could be obtained by
 \be
{\cal B}(\bar B_d\to \pi^{+} \pi^{-} )&=& \frac{\tau_{B_d}}{16\pi
m_B}|{\cal M}^{\UP}_{\pi^+ \pi^-}|^2\,, \non \\
{\cal A}_{CP}&=& \frac{\bar {\cal B}(\bar B_d\to \pi^{+} \pi^{-} )-
{\cal B}( B_d\to \pi^{-} \pi^{+} )}{\bar {\cal B}(\bar B_d\to
\pi^{+} \pi^{-} ) + {\cal B}( B_d\to \pi^{-} \pi^{+} )}
\label{eq:cp-acp}
 \ed
where $\tau_{B_d}$ is the lifetime of $B_d$ and the pion mass has
been neglected. To be more clear to see the relationship of the CPA
with the unparticle
phase $\phi_{\UP}$, we rewrite the CPA for $\bar B_d\to \pi^{+}
\pi^{-}$ as
 \be
{\cal A}_{CP}=\frac{-2 \chi^{\UP}_{\pi\pi}\sin \alpha
\sin\phi_{\UP}}{|1+\chi^{SM}_{\pi\pi} e^{i\alpha}|^2 +
|\chi^{\UP}_{\pi\pi}|^2 -2\chi^{\UP}_{\pi\pi} \cos\alpha
\cos\phi_{\UP}}
 \ed
where $\alpha\equiv \beta + \ga$ and
 \be
 \chi^{\UP}_{\pi\pi}&=& \frac{\tilde{\Delta}_{\UP}(p^2)}{4N_c a_1} \frac{|\xi^{d}_{t}|}{|\xi^{d}_{u}|}
 \left(C^{u}_{L} +2 r_{\pi} C^{u}_{R} \right)\,,\non\\
  \chi^{SM}_{\pi\pi}&=& \frac{1}{a_1} \frac{|\xi^{d}_{t}|}{|\xi^{d}_{u}|}
 \left(a^{u}_{4} +2 r_{\pi} a^{u}_{6} \right)\,.
   \ed
Note that
$a^{q}_{4(6)}$ can be derived from Eq.~(\ref{eq:effwcs}) by
setting $C^{q}_{L(R)}=0$. Obviously, the CPA in $\bar
B_d\to \pi^{+} \pi^{-}$ depends on not only
the weak phase $\alpha$ but also the CP-conserved unparticle phase $\phi_{\UP}$.

\subsection{$B\to \pi K$  decays }

  It is known that the decays of $b\to s q\bar q$ are
penguin dominant processes as the tree contributions
are suppressed by the CKM matrix elements of
$V_{ub} V^*_{us}$.
 Since the unparticle effects are also
induced from penguin loops, one  expects that they should be
significant. In $B\to \pi K$ decays, there are four specific decay
modes. Since the BRs for $B\to \pi K$ and CPA of $B_d\to \pi^{-}
K^{+}$ are observed well in experiments, we have to discuss all
modes in detail. We begin with our analysis on the decay of
$B^{-}\to \pi^{-} \bar K^{0}$.
%%%%%%%%%%%%%%%%%%%%%%%%%%%%%%%%%%%%%%%%%%%%%%%%%%%%%%%%%%%%%
\begin{figure}[hpbt]
\includegraphics*[width=4. in]{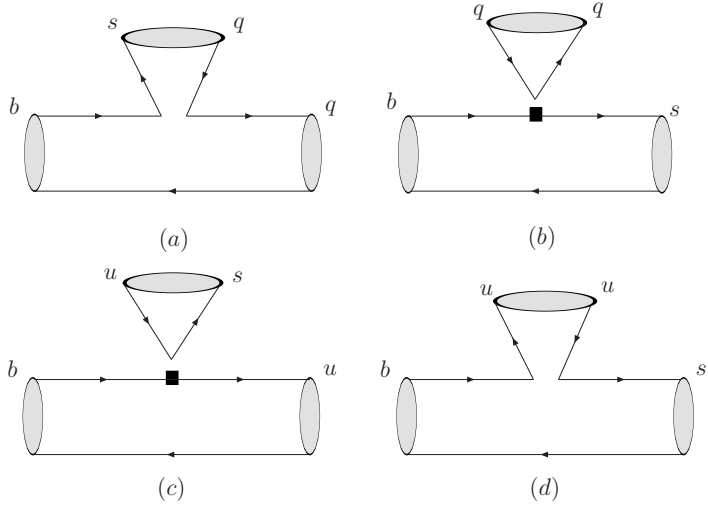}
\caption{Flavor diagrams for $B\to \pi K$ decays dictated by
(a)[(b)] penguin and (c)[(d)] tree diagrams.}
 \label{fig:bpik}
\end{figure}
%%%%%%%%%%%%%%%%%%%%%%%%%%%%%%%%%%%%%%%%%%%%%%%%%%%%%%%%%%%%
According to the flavor diagram in Fig.~\ref{fig:bpik}a, the decay
is corresponding to $q=d$. Hence, taking the same conditions as $B_d\to
\pi^{-} \pi^{+}$, the decay amplitude for $B^{-} \to \pi^{-} \bar
K^{0} $ can be expressed by
\begin{eqnarray}
{\cal M}^{\UP}_{\pi^- \bar K^0}&=& - \frac{G_F}{\sqrt{2}}
V_{tb}V^{*}_{ts}f_{K} m^2_{B} F^{B\pi}_0 (a^{d\UP}_4 + 2 r_{K}
a^{d\UP}_6)\,, \label{eq:amp_cpink}
\end{eqnarray}
where $f_{K}$ is the kaon decay constant and
$r_{K}=m^{0}_{K}/m_{B}$
with $m^{0}_{K}$ defined as $m^{0}_{\pi}$.
 Similar to
$B^{-}\to \pi^{-} \bar K^0$, we can easily find the decay amplitude of
$\bar B_{d}\to \pi^{+} K^-$ by using $q=u$ instead of $q=d$,
given by
\begin{eqnarray}
{\cal M}^{\UP}_{\pi^+ K^-}&=& \frac{G_F}{\sqrt{2}} f_K m^2_B
F^{B\pi}_{0 } \left[-V_{tb}V^{*}_{ts}  \left( a_4^{u\UP} + 2  r_{K}
a^{u\UP}_6 \right)
 + V_{ub} V^*_{us}  a_{1}\right], \label{eq:amp_cpick}
\end{eqnarray}
where we have included the tree contributions  illustrated in
Fig.~\ref{fig:bpik}c.

Next, we analyze the decay of $B_{d}\to\pi^{0} K^0$.
Besides the flavor diagrams appearing
in the decay of $B^{-}\to \pi^{-} \bar K^0$, there are  new diagrams
shown in Figs.~\ref{fig:bpik}b and \ref{fig:bpik}d. In the SM, these
contributions correspond to the electroweak penguin and
color-suppressed effects, respectively. Taking $q=u$ and $d$ in
Figs.~\ref{fig:bpik}a and
\ref{fig:bpik}b, respectively, the
decay amplitude for $\bar B_d\to \pi^{0} \bar K^0$ is given by
\begin{eqnarray}
\sqrt{2}{\cal M}^{\UP}_{\pi^0 \bar K^0}&=&\frac{G_F}{\sqrt{2}} m^2_B
\left\{V_{tb}V^{*}_{ts} \left[f_{K}  F^{B\pi}_{0 } ( a^{d\UP}_4+ 2
r_{K} a^{d\UP}_6 )- f_{\pi} \zeta F^{BK}_0 \right] + V_{ub} V^*_{us}
f_{\pi} a_{2} F^{BK}_{0} \right\}, \label{eq:amp_npink}
\end{eqnarray}
where $\zeta=a^{u\UP}_{3}-a^{d\UP}_{3}+a^{d\UP}_{5}-a^{u\UP}_{5} $.
We note that the new term $f_{\pi} \zeta F^{BK}_0$, corresponding to
the contribution in
Fig.~\ref{fig:bpik}b, has opposite in sign to other terms. The
reason comes from the flavor wave function of $\pi^{0}$ being
$(\bar{u} u-\bar{d} d)/\sqrt{2}$. Note that Fig.~\ref{fig:bpik}b picks
both components while Fig.~\ref{fig:bpik}a only takes the $\bar{d} d$ component. Since
the tree contributions are color suppressed, the corresponding WC is
$a_{2}$.

After introducing the decay amplitudes for $B^{-}\to \pi^{-} \bar
K^0$, $\bar B_d\to \pi^{+} K^-$ and $\bar B_d\to \pi^0 \bar K^0 $,
the amplitude of $B^-\to \pi^{0} K^-$  could be immediately
obtained as
\begin{eqnarray}
\sqrt{2}{\cal M}^{\UP}_{\pi^0K^-}&=&\frac{G_F}{\sqrt{2}} m^2_B
\left[V_{tb} V^{*}_{ts} \left(-  f_{K} F^{B\pi}_{0 } ( a^{u\UP}_4 +
2 r_{K} a^{u\UP}_6 ) - f_{\pi} \zeta
F^{BK}_{0}\right) \right. \nonumber \\
 && \left. + V_{ub} V^*_{us}  (f_{K} a_1 F^{B\pi}_0+ f_{\pi} a_{2} F^{BK}_{0})\right].
 \label{eq:amp_npick}
\end{eqnarray}
Clearly, the amplitudes for the first three decay modes all
appear in the decay of $B^+\to \pi^0 K^+$. That is,
once the first three decays are determined, the decay of $B^{+}\to \pi^0 K^+$ is also
fixed. We would point out that although Fig.~\ref{fig:bpik}b
 contributes to the modes of $\pi^{0} \bar K^0$ and $\pi^0 K^{+}$,
 similar to the case in $B\to \pi \pi$,  the unparticle
effects in this topology will vanish
if we take
 $C^{d}_{L(R)} \approx C^{u}_{L(R)}$. In our study, we also
neglect their contributions. In sum,
the BRs and CPAs for the $B\to \pi K$ decays can be
found by  the definitions in Eq.~(\ref{eq:cp-acp}).

\subsection{ Inclusive semileptonic decays of $b\to q \bar \ell \ell$}

\begin{figure}[hpbt]
\includegraphics*[width=2.5 in]{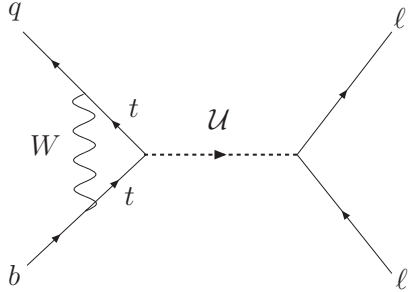}
\caption{$b\to q \ell^{+} \ell^{-}$ decays induced by unparticle
penguin diagram.}
 \label{fig:up_bqll}
\end{figure}

If unparticles couple to leptons, we can apply the induced
interactions for $b\to q\UP$ to study the semileptonic decays of
$b\to q \bar \ell \ell$. The corresponding Feynman diagram is presented in
Fig.~\ref{fig:up_bqll}. It is easy to see that due to the CKM
suppression,  the semileptonic decays with $b\to d$ are much less than those
of $b\to s$.
Hence, in the following discussions, we will
concentrate on $b\to s \ell^{+} \ell^{-}$. Nevertheless, all
discussions and formulas could be applied to $b\to d \ell^{+}
\ell^{-}$ as well. It is also worth mentioning that because the CKM matrix
element $V_{td}$ carries a CP violating phase, the system of $b\to d
\ell^{+} \ell^{-}$ could be even more interesting on CP violation in
the framework of unparticle physics.

%%%%%%%%%%%%%%%%%%%%%%%%%%%%%%%%%%%%%%%%%%%%%%%%%%%%%%%%%%%%%
%%%%%%%%%%%%%%%%%%%%%%%%%%%%%%%%%%%%%%%%%%%%%%%%%%%%%%%%%%%%

%
Before including new physics interactions into $b\to s \ell^{+}
\ell^{-}$, we write the effective Hamiltonian for the SM as
  \be
 {\cal H}_{\rm eff}= \frac{G_F\alpha_{em} \lambda_t}{\sqrt{2}
 \pi}\left[ H_{1\mu} L^{\mu} +H_{2\mu}L^{5\mu}  \right]
 \label{eq:hbqll}
  \en
  with
  \begin{eqnarray}
  H_{1\mu } &=&C^{\rm eff}_{9}(\mu )\bar{s}\gamma _{\mu }P_{L}b\ -\frac{2m_{b}}{%
 q^{2}}C_{7}(\mu )\bar{s}i\sigma _{\mu \nu }q^{\nu }P_{R}b \,,
\nonumber \\
 H_{2\mu } &=&C_{10}\bar{s}\gamma _{\mu }P_{L}b \,,
 \nonumber\\
 L^{\mu } &=&\bar{\ell}\gamma ^{\mu }\ell\,, \ \ \ L^{5\mu } =\bar{\ell}\gamma ^{\mu }\gamma
 _{5}\ell\,,
  \label{eq:heffc}
  \end{eqnarray}
where $\alpha_{em}$ is the fine structure constant,
$\lambda_t=V_{tb}V_{ts}^*$,  $C^{\rm eff}_{9}$ and $C_{7,10}$ are
the Wilson coefficients (WCs) with their explicit expressions given
in Ref.~\cite{BBL} for the SM, $m_b$ is the current b-quark mass and
$q^2$ is the invariant mass of the $\ell^{+}\ell^{-}$ pair. Although
long-distance effects of $c\bar{c}$ bound states could contribute to
$C^{\rm eff}_9$, to study the behavior of unparticle physics in the
semileptonic decays, for simplicity they are not included in the
present study. On the other hand, the bound states could be excluded
experimentally by cutting the phase space at the resonant regions.
Explicitly, one has that \cite{BBL}
\begin{eqnarray}
C_{9}^{\rm eff}(\mu)&=&C_{9}( \mu ) +\left( 3C_{1}\left( \mu \right)
+C_{2}\left( \mu \right) \right) h\left( x,s\right) \,, \nonumber \\
h(z,s)&=&-\frac{8}{9}\ln\frac{m_b}{\mu}-\frac{8}{9}\ln z
+\frac{8}{27} +\frac{4}{9}x  -\frac{2}{9}(2+x)|1-x|^{1/2} \nonumber
\\
&\times& \left\{
  \begin{array}{c}
    \ln \left|\frac{\sqrt{1-x}+1}{\sqrt{1-x}-1} \right|-i\, \pi, \  {\rm for\ x\equiv 4z^2/s<1 }\, , \\
    2\, arctan\frac{1}{\sqrt{x-1}},\   {\rm for\ x\equiv 4z^2/s>1 }  \, ,\\
  \end{array}
\right. \label{eq:onshell}
\end{eqnarray}
where $h(z,s)$ describes the one-loop matrix elements of operators
$O^{c}_{1}= \bar{s}_{\alpha }\gamma ^{\mu }P_{L}b_{\beta }\
\bar{c}_{\beta }\gamma _{\mu }P_{L}c_{\alpha }$ and
$O^{c}_{2}=\bar{s}\gamma ^{\mu }P_{L}b\ \bar{c}\gamma _{\mu }P_{L}c$
\cite{BBL} with $z=m_c/m_b$ and $s=q^2/m^2_b$. Comparing to
Eq.~(\ref{eq:h_U}), we find that the operator structures of the unparticle
contributions are the same as those of the SM. The unparticle effects
with the SM contributions can be derived by using $C^{\UP}_9$ and
$C^{\UP}_{10}$, defined by
 \be
 C^{\UP}_{9}(q^2) &=& C^{\rm eff}_9 + \frac{\pi}{\alpha_{em}}
\frac{C^{\ell}_R + C^{\ell}_L}{2}\tilde{\Delta}_{\UP}(q^2) e^{-i
\phi_{\UP}}\,, \non\\
 C^{\UP}_{10}(q^2) &=& C_{10}+ \frac{\pi}{\alpha_{em}}
\frac{C^{\ell}_R - C^{\ell}_L}{2} \tilde{\Delta}_{\UP}(q^2) e^{-i
\phi_{\UP}}\,, \label{eq:c910}
 \ed
instead of $C^{\rm eff}_9$ and $C_{10}$, respectively.

With Eq.~(\ref{eq:hbqll}) and  the three-body phase
space, the inclusive differential decay rate for $b\to s \ell^{+}
\ell^{-}$ ($\ell=e\,, \mu)$ can be expressed by
  \be
  \frac{d\Gamma}{ ds}&=& \frac{G^2_F m^5_b
\alpha^2_{\rm
em}}{768\pi^5} |V_{ts} V^{*}_{tb}|^2(1-s)^2 R(s)\,,\non\\
R(s)&=& \left( |C^{\UP}_{9}(s)|^2+ |C^{\UP}_{10}(s)|^2\right) (1+2
s)+12 Re(C^{*}_{7} C^{\UP}_9(s)) + 4\left(
1+\frac{2}{s}\right)|C_7|^2\,, \label{eq:ratecull}
 \ed
where the lepton mass has been neglected. Besides the BRs, it has
been known that the forward-backward asymmetry (FBA), defined by
\cite{FBA}
 \be
 \frac{d A_{FB}}{ ds}&=& {\int ^{1}_{-1} d\cos\theta
 d\Gamma/ds\,d\cos\theta
 \, {\rm sgn(\cos\theta)}\over \int^{1}_{-1} d\cos\theta
 d\Gamma/ds\,d\cos\theta }\,,
 \label{AFB}
  \ed
could be as a good candidate to probe new physics.
By including the unparticle contributions,
from Eqs. (\ref{eq:ratecull}) and (\ref{AFB}) we get
 \be
 \frac{d A_{FB}}{ ds}
&=&  -3 \frac{s}{R(s)} Re\left[ \left(C^{\UP}_9(s) + \frac{2}{s}
C_7\right) C^{\UP^*}_{10}(s)\right]\,. \label{eq:fba}
  \ed
We note that as a whole the decay of $b\to s \ell^{+} \ell^{-}$ is
not sensitive to the CP-conserved phases carried by the unparticle
propagators. However, it is important to point out that at the low
$q^2$ regions the unparticle physics has large effects on these
physical quantities as the unparticle propagator is proportional to
$(q^2)^{d_{\UP} -2}$. In these region, we expect that  both BR and
FBA in $b\to s \ell^{+} \ell^{-}$ have significant deviations from
the SM predictions.

\section{Numerical analysis}{\label{sec:NA}}

To estimate the numerical values, we take the common
parameters to be: $G_F=1.166\times 10^{-5}$ GeV$^{-2}$,
$f_{\pi}=0.13$ GeV, $f_{K}=0.16$ GeV, $m^0_K=2.4$ GeV,
$m^0_{\pi}=1.7$ GeV, $V_{td}=8.46\times 10^{-3} e^{-i\beta}$ with
$\beta=25^{\circ}$, $V_{ub}=3.6\times 10^{-3} e^{-i \ga}$ with
$\ga=72^{\circ}$, and $\alpha_{\rm em}=1/129$ \cite{PDG06}. For the
nonleptonic B decays, since we concentrate on the CPAs,
the CP-averaged BRs are regarded as inputs and their world averages
are adopted as \cite{HFAG}
 \be
  {\cal B}(\bar B_d\to \pi^+ \pi^-)&=&(5.16\pm 0.22)\times 10^{-6}\,, \non\\
  {\cal B}(B^-\to \pi^- \bar K^0)&=&(23.1\pm 1.0)\times 10^{-6}\,, \non\\
  {\cal B}(\bar B_d\to \pi^+ K^-)&=&(19.4 \pm 0.6)\times 10^{-6}\,, \non\\
   {\cal B}(\bar B_d\to \pi^0 \bar K^0)&=&(10.0 \pm 0.6)\times 10^{-6}\,, \non\\
  {\cal B}(B^- \to \pi^0 K^-)&=&(12.8 \pm 0.6)\times 10^{-6}\,.
 \ed
For the semileptonic $b\to s \ell^{+} \ell^{-}$ decays, we use the world
average as \cite{PDG06}
 \be
 {\cal B}(b\to s \ell^{+} \ell^{-})&=& (4.5 \pm 1.0)\times
 10^{-6}\,.
 \ed

\begin{figure}[hpbt]
\includegraphics*[width=4. in]{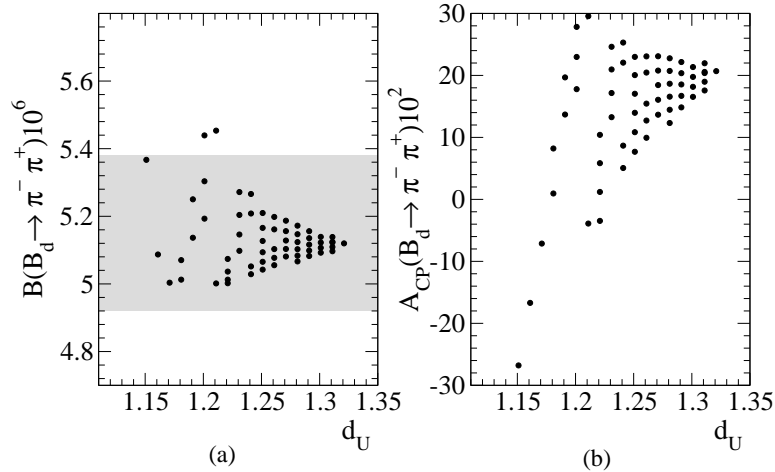}
\caption{(a) BR (in units of $10^{-6}$) and (b) CPA ($\%$) for
$B_d\to \pi^{-} \pi^{+}$ versus the scale dimension $d_{\UP}$ with
$\Lambda_{\UP}=1$ TeV and $\lam^{q}_{L(R)}< 0.5$, where the band in
(a) denotes the world average with $1\sigma$ errors.}
 \label{fig:brcp_bpipi}
\end{figure}

 We first calculate the unparticle contributions to the BR and CPA of
$\bar B_d\to \pi^{+} \pi^{-}$. As mentioned before, to satisfy the
indication of the experimental data on $B^- \to \pi^{0} \pi^{-}$, we
will require that $C^{u}_{L(R)}\approx C^{d}_{L(R)}=C^{q}_{L(R)}$.
Besides the scale dimension $d_{\UP}$ and the unparticle scale
$\Lambda_{\UP}$ fixed to be 1 TeV, there are three extra unknown
parameters from unparticle physics, {\it i.e.} $C^t_{L}$, $C^{q}_L$
and $C^{q}_R$. Since $C^{t}_{L}$ is always associated with
$C^{q}_{L(R)}$, in our numerical calculations, we will use the
combined parameters of $\lam^{q}_L=C^t_L C^{q}_L$ and
$\lam^{q}_{R}=C^{t}_L C^{q}_R$. In addition, we will set the
available ranges for the variables to be $|\lam^{q}_{L(R)}|< 0.5$.
In terms of Eqs.~(\ref{eq:amp_cpicpi}) and (\ref{eq:cp-acp}), the
numerical values of the BR and CPA for $\bar B_d \to \pi^{-}
\pi^{+}$ versus the scale dimension $d_{\UP}$ are presented in
Fig.~\ref{fig:brcp_bpipi}. Because the BR is input, the calculated
values are all within $1\sigma$ errors. Recently, BABAR \cite{babar}
and BELLE \cite{belle} have reported the CPA measurements of
 \be
  {\cal A}_{CP}(\bar B_d\to \pi^+ \pi^{-})=(0.21 \pm 0.09 \pm 0.02)\ \ \
  {\rm (BABAR)}\,, \non\\
  {\cal A}_{CP}(\bar B_d\to \pi^+ \pi^{-})=(0.55 \pm 0.08 \pm 0.05)\ \ \
  {\rm (BELLE)}\,,
 \ed
with the average value being
\be
{\cal A}_{CP}(\bar B_d\to \pi^+ \pi^{-})&=&0.38 \pm 0.18.
\ed
 According to our
results in Fig.~\ref{fig:brcp_bpipi}b, we see  that without
any QCD phases, the unparticle-mediated FCNC with
the peculiar CP-conserved phase induced by
the penguin diagram
could make the CPA of
$\bar B_d \to \pi^{+} \pi^{-}$ as large as $20\%$.
Clearly, with more and more data
accumulated at the B factories, it is worth to explore
whether the unparticle phase is the dominant source for
dictating the CPA of $\bar B\to \pi^{+} \pi^{-}$.
%%%%%%%%%%%%%%%%%%%%%%%%%%%%%%%%%%%%%%%%%%%%%%%%%%%%%%%%%%%%%
%%%%%%%%%%%%%%%%%%%%%%%%%%%%%%%%%%%%%%%%%%%%%%%%%%%%%%%%%%%%

\begin{figure}[hpbt]
\includegraphics*[width=4.5 in]{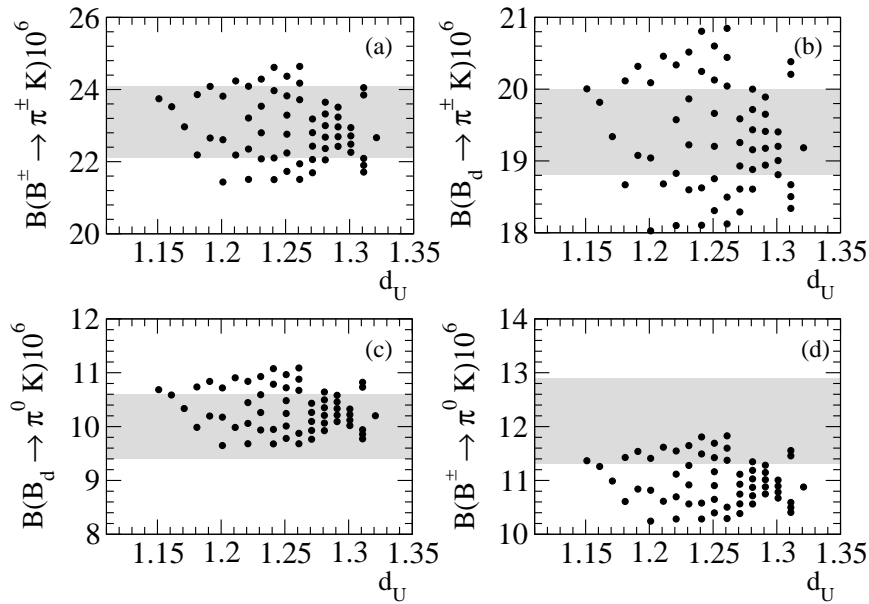}
\caption{ BRs (in units of $10^{-6}$) versus the
scale dimension $d_{\UP}$ for (a) $B^- \to \pi^- \bar K^0$, (b) $\bar B_d
\to \pi^0 \bar K^0$, (c) $\bar B_d \to \pi^0 \bar K^0$ and (d) $B^-\to \pi^0 K^-$, where the band in the each figure stands for
the world average with $1\sigma$ errors.}
 \label{fig:br_bpiK}
\end{figure}

With the same set of parameters
in $B\to
\pi\pi$, we now study the decays of $B\to \pi K$. According to the
formulas of the decay amplitudes introduced in
Eqs.~(\ref{eq:amp_cpink}), (\ref{eq:amp_cpick}),
(\ref{eq:amp_npink}) and (\ref{eq:amp_npick}), the values of BRs
through unparticle-mediated diagrams are presented in
Fig.~\ref{fig:br_bpiK}, where the band in each figure denotes the
world average with $1\sigma$ errors.
It is interesting to see that
unparticle physics could
make the BRs of $B\to \pi K$ be consistent with data within
$1\sigma$ world averages. We note that the data of ${\cal
B}(\bar B_d \to \pi^{+} \pi^{-})$ has been also included to
constrain the various unknown parameters.

Although the consistent results in the BRs have been impressive
enough, to emphasize the importance of the magic phase in unparticle
physics, one should pay attention to the CPA. According to our
previous analysis in Eqs.~(\ref{eq:amp_cpink}) and
(\ref{eq:amp_npink}), since the tree contributions are negligible
(small) for $B^{-} \to \pi^- \bar K$ ($\bar B_d \to \pi^0 \bar
K^0$), one can easily understand that the corresponding CPA should
be also negligible (small). Nevertheless, we have to remark that the
conclusions are correct only for the cases without including final
state interactions (FSIs). Note that we have to exclude the
discussions on FSIs since we can only control the short-distance
effects. Here, we adopt that the assumption of color transparency
dominates the processes in $B$ decays \cite{Bjorken}.
We conclude that the interesting CP violating effects
in $B\to \pi K$ decays are the CPAs for $\bar B_d\to \pi^{+} K^-$
and $B^-\to \pi^0 K^-$.
%%%%%%%%%%%%%%%%%%%%%%%%%%%%%%%%%%%%%%%%%%%%%%%%%%%%%%%%%%%%%
%%%%%%%%%%%%%%%%%%%%%%%%%%%%%%%%%%%%%%%%%%%%%%%%%%%%%%%%%%%%

\begin{figure}[hpbt]
\includegraphics*[width=4. in]{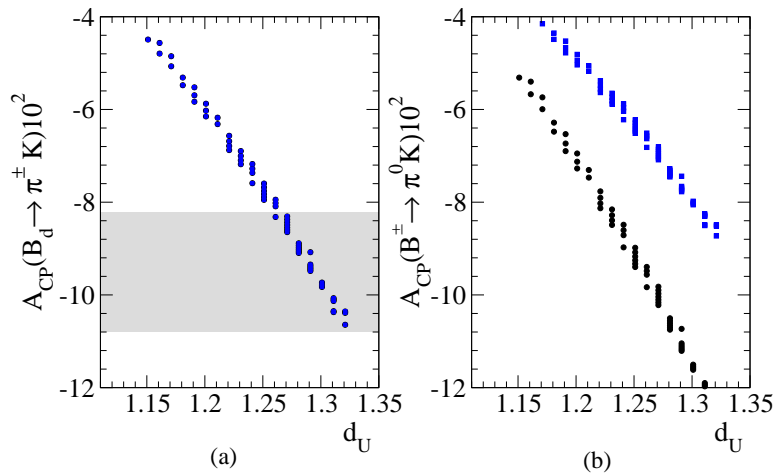}
\caption{CPAs ($\%$) for (a) $B_d\to \pi^{\mp} K^{\pm}$ and (b)
$B^{\mp} \to \pi^{0} K^{\mp}$ versus the scale dimension $d_{\UP}$,
where the band in (a) denotes the world average with $1\sigma$
errors and the circles (squares) dots in each figure represent
$a_2=0.14 (-0.14)$.}
 \label{fig:acp_bpiK}
\end{figure}

 From Eqs.~(\ref{eq:amp_cpick}) and (\ref{eq:amp_npick}), one
finds that the penguin contributions are the same in both decays,
the only difference is that there is an extra color-suppressed
contribution in $B^- \to \pi^0 K^-$. If the color-suppressed $a_2$
term in Eq.~(\ref{eq:amp_npick}) is dropped, one expects that both
modes should have the same CPAs. Therefore, the sign of $a_2$ will
affect the CPA of $B^-\to \pi^0 K^-$. Using the parameters fitted by
the BRs of $B\to \pi K$, we present the unparticle contributions
versus the scale dimension $d_{\UP}$ in Fig.~\ref{fig:acp_bpiK},
where the circles (squares) dot in each figure denotes $a_2=0.14
(-0.14)$. We see  that the sign of $a_2$ has a significant influence
on ${\cal A}_{CP}(B^-\to \pi^0 K^-)$.  According to the current
world average, given by \cite{HFAG}
 \be
  {\cal A}_{CP}(\bar B_d \to \pi^+ K^-)= -0.095 \pm 0.013\,, \non\\
  {\cal A}_{CP}(B^- \to \pi^0 K^-)= 0.047 \pm 0.026\,,
  \label{eq:data_cp}
 \ed
it seems that somewhat different physics exists between the two
modes. Plausibly, $a_2$ plays an important role in the CPA for
$B^{-} \to \pi^0 K^-$. Since our focus is on the CPA in unparticle
physics, further discussions on $a_2$ are given elsewhere. The
detailed analysis could refer to Ref.~\cite{PQCD}.
 From our results, we see that with the unparticle phase,
${\cal A}_{CP}(B^{-}\to \pi^{+} K^-)$ could be consistent with the
current data. As the results in Eq.~(\ref{eq:data_cp}) are
not conclusive yet,
more precise data are needed to tell whether there
is a deviation between $\bar B_d\to \pi^{-} K^+$ and $B^{-}\to \pi^0
K^-$ in the CPAs.
%%%%%%%%%%%%%%%%%%%%%%%%%%%%%%%%%%%%%%%%%%%%%%%%%%%%%%%%%%%%%
%%%%%%%%%%%%%%%%%%%%%%%%%%%%%%%%%%%%%%%%%%%%%%%%%%%%%%%%%%%%

Finally, we study the unparticle effects on inclusive semileptonic
decays of $b\to s \ell^{+} \ell^{-}$ with $\ell=e\,, \mu$. From
Eq.~(\ref{eq:c910}), one finds that $C^{t}_{L}$ is always associated
with $(C^{\ell}_R + C^{\ell}_{R})/2$ and $(C^{\ell}_R -
C^{\ell}_{L})/2$. To simplify our numerical analysis, we set
$C^{\ell}_R= C^{\ell}_L$ or $C^{\ell}_R= -C^{\ell}_L$. We will
redefine our parameters to be $C^t_L C^{\ell}_L=C^t_L
C^{\ell}_R=\lam^{\ell}_V$ and $C^t_L C^{\ell}_L=-C^t_L
C^{\ell}_R=\lam^{\ell}_A$ and discuss the constraints on
$\lam^{\ell}_{V(A)}$.
  From Eq.~(\ref{eq:c910}), we
know that the one-loop matrix elements from operators $O^{c}_1$ and
$O^{c}_2$ will generate a CP-conserved QCD phase, which in principle
could  interfere with the unparticle phase. However, the
interference effect between the $CP$-conserved phases of the SM and
unparticles is small since the one-loop generated contributions are
much smaller than $C_9\sim -C_{10}\sim 4$.

\begin{figure}[hpbt]
\includegraphics*[width=4. in]{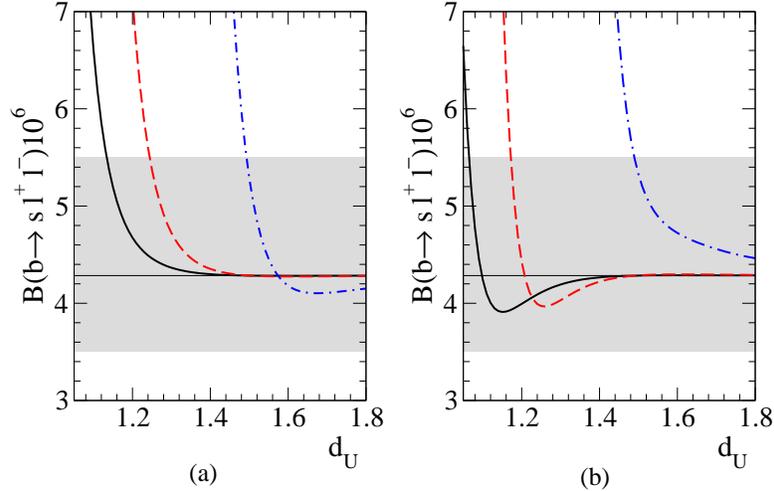}
\caption{BR ( in units of $10^{-6}$) for $b\to s \ell^{+} \ell^{-}$
with (a) $C^t_{L} C^{\ell}_R = C^t_L C^{\ell}_{L}=\lam^{\ell}_V$ and
(b) $C^t_{L} C^{\ell}_R =-C^t_L C^{\ell}_{L}=\lam^{\ell}_{A}$ versus
the scale dimension $d_{\UP}$, where the thick solid, dashed and
dash-dotted lines correspond to $\lam^{\ell}_{V(A)}=0.005$, $0.01$
and $0.05$, respectively, while the horizontal  lines are the SM
contributions and the bands denote the world average with $1\sigma$
errors.}
 \label{fig:br_bsll}
\end{figure}

Although $b\to s \ell^{+} \ell^{-}$ cannot be the candidate to probe
the unique unparticle phase,  we can utilize the decays to give
strong constraints on the unparticle couplings to leptons, {\it
i.e.} $\lam^{\ell}_{V(A)}$. In terms of Eq.~(\ref{eq:ratecull}) and
the values for the common parameters, the unparticle contributions
to the BRs of $b\to s \ell^{+} \ell^{-}$ are presented in
Fig.~\ref{fig:br_bsll}, where
%Fig.~\ref{fig:br_bsll}
(a)[(b)] denotes the contributions of $\lam^{\ell}_{V[A]}$, the
horizontal thin lines are the SM contributions and the thick solid,
dashed and dash-dotted lines correspond to
$\lam^{\ell}_{V[A]}=0.005$, $0.01$ and $0.05$, respectively. The
bands in the diagrams are the world average with $1\sigma$ errors.
 From the figure, we see clearly that with a specific value for
$\lam^{\ell}_{V(A)}$, the BR of $b\to s \ell^{+} \ell^{-}$ is very
sensitive to the scale dimension $d_{\UP}$.
%%%%%%%%%%%%%%%%%%%%%%%%%%%%%%%%%%%%%%%%%%%%%%%%%%%%%%%%%%%%%
%%%%%%%%%%%%%%%%%%%%%%%%%%%%%%%%%%%%%%%%%%%%%%%%%%%%%%%%%%%%
To understand the sensitivity, we need to examine the behavior of
the unparticle propagator  and the unparticle couplings to
fermions. With Eqs.~(\ref{eq:lang_UP}) and (\ref{eq:uprop}), we know
that the $q^2$-dependence in the BR will behave like
 \be
 \left[\frac{1}{q^2}\left(\frac{q^2}{\Lambda^2_{\UP}}\right)^{d_{\UP}-1}\right]^{2}\,.
  \ed
It is clear that for the three-body $b\to s \ell^{+} \ell^{-}$ decays,
the enhancement of unparticle effects is  at the small invariant mass of
$\ell^{+} \ell^{-}$.
%%%%%%%%%%%%%%%%%%%%%%%%%%%%%%%%%%%%%%%%%%%%%%%%%%%%%%%%%%%%%
\begin{figure}[hpbt]
\includegraphics*[width=4. in]{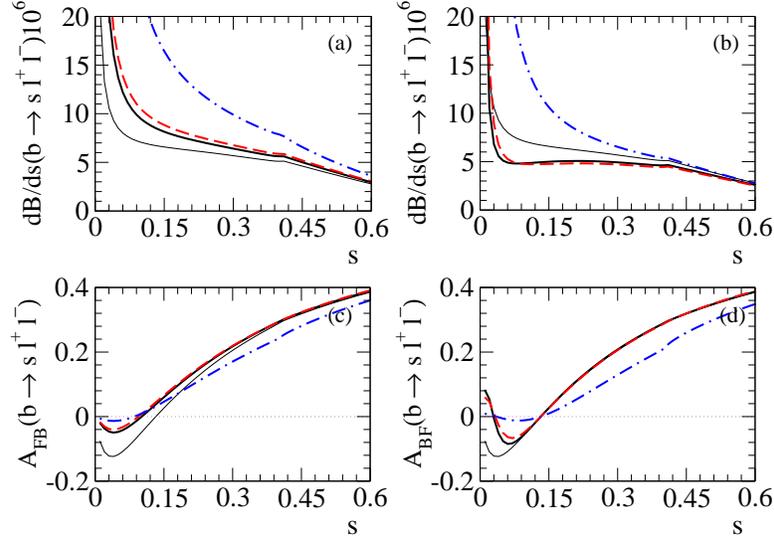}
\caption{ (a)[(b)] Differential BR ( in units of $10^{-6}$) and
(c)[(d)] FBA for $b\to s \ell^{+} \ell^{-}$ as  functions of
$s=q^2/m^2_b$, where the thick solid, dashed and dash-dotted lines
correspond to $d_{\UP}(\lam^{\ell}_{V[A]})=1.1 (0.005)$, $1.2
(0.01)$ and $1.4 (0.05)$, respectively, while the thin lines are the
SM contributions. }
 \label{fig:difbr-fba}
\end{figure}
%%%%%%%%%%%%%%%%%%%%%%%%%%%%%%%%%%%%%%%%%%%%%%%%%%%%%%%%%%%%
To be more clear, we display the differential BRs for $b\to s
\ell^{+} \ell^{-}$ as  functions of $s=q^2/m^2_b$ in
Figs.~\ref{fig:difbr-fba}a and \ref{fig:difbr-fba}b, where the
solid, dashed and dash-dotted lines stand for
$d_{\UP}[\lam^{\ell}_{V(A)}]=1.1 [0.005],\, 1.2 [0.01]$ and $1.4
[0.05]$. The large deviation at the small $s$ region could confirm
our argument.

 For the FBA in $b\to s\ell^{+}
\ell^{-}$,
 from Eq.~(\ref{eq:fba}), the numerical
values of the unparticle contributions as  functions of the
invariant mass $s$ are shown in Figs.~\ref{fig:difbr-fba}c and
\ref{fig:difbr-fba}d. Clearly, the FBA at the small $s$ region  is
also sensitive to the unparticle physics. In addition, one observes
that the nonvanished $\lam^{\ell}_{V}$ associated with $C^{\UP}_9$
could shift the zero point of the FBA to be lower. However, the
nonvanished $\lam^{\ell}_{A}$ associated with $C^{\UP}_{10}$ cannot
change the zero-crossing point. The reason could be understood from
Eq.~(\ref{eq:fba}), where the zero point can only happen at
$C^{\UP}_9+2 C_7/s=0$.

\section{Conclusions}\label{sec:conclusion}
We have studied the implications of the CP conserving phases in the
unparticle propagators. We have demonstrated that these peculiar
phases have an important impact on CP violation
since they could act as the strong phases needed to induce the direct
CP asymmetry.
Without including the QCD phases,
we have examined the unparticle phase effects on the
direct CP asymmetries in the
exclusive $\bar B_d\to \pi^{+} \pi^{-}$ and $B\to \pi K$ decays,
in which FCNCs are forbidden at tree level but induced by one-loop unparticle penguin diagrams.
We have obtained interesting and
consistent results comparing to the experimental data.
Moreover, we have found that the unparticle effects will significantly enhance
the differential branching ratio of  $b\to s \ell^{+}
\ell^{-}$ at the
small invariant mass of $\ell^{+} \ell^{-}$
so that the couplings of unparticles to leptons suffer
strong constraints. The forward-backward
asymmetries for the decays of $b\to s \ell^{+} \ell^{-}$ due to
the unparticle effects have also been investigated.\\

\noindent
{\bf Acknowledgments}

 This work is supported in part by the National
Science Council of R.O.C. under Grant \#s: NSC-95-2112-M-006-013-MY2
and NSC-95-2112-M-007-059-MY3.

%%%%%%%%%%%%%%%%%%%%%%%%%%%%%%%%%%%%%%%%%%%%%%%%%%%%%%

\end{document}